\UseRawInputEncoding
\documentclass[showpacs,twocolumn,superscriptaddress,nofootinbib]{revtex4-1} 

\usepackage{hyperref}
\hypersetup{colorlinks=true,linkcolor=blue,citecolor=cyan}
\usepackage{graphicx}
\usepackage{xcolor, soul}
\sethlcolor{green}
\usepackage{dcolumn}
\usepackage{bm}
\usepackage{color}
\usepackage{enumitem}
\usepackage{mathrsfs}
\usepackage{amsmath}
\usepackage{amssymb}
\usepackage{cleveref}
\usepackage{orcidlink}

\crefname{equation}{Eqn.}{Eqns.}
\crefname{figure}{Fig.}{Figs.}
\crefname{section}{Sec.}{Sec.}
\crefname{table}{Table}{Tables}
\usepackage{physics}
\usepackage{multirow}
\usepackage{diagbox}
\setstcolor{red}

\newcommand{\cw}[1]{\ignorespaces}

\begin{document}

\title{Non-Monotonic Enhancement of the Magnetic Penrose Process in Kerr-Bertotti-Robinson Spacetime and its Implication for Electron Acceleration
}

\author{Mirjavoxir Mirkhaydarov}
\email{mirxaydarovmirjavohir@gmail.com}
\affiliation{National University of Uzbekistan, Tashkent 100174, Uzbekistan}

\author{Tursunali Xamidov
}
\email{xamidovtursunali@gmail.com}
\affiliation{Institute of Fundamental and Applied Research, National Research University TIIAME, Kori Niyoziy 39, Tashkent 100000, Uzbekistan}
\affiliation{Institute for Theoretical Physics and Cosmology, Zhejiang University of Technology, Hangzhou 310023, China}

\author{Pankaj Sheoran
}
\email{pankaj.sheoran@vit.ac.in}
\affiliation{Department of Physics, School of Advanced Sciences, Vellore Institute of Technology, Tiruvalam Rd, Katpadi, Vellore, Tamil Nadu 632014, India}

\author{Sanjar Shaymatov
}
\email{sanjar@astrin.uz}
\affiliation{Institute of Fundamental and Applied Research, National Research University TIIAME, Kori Niyoziy 39, Tashkent 100000, Uzbekistan}
\affiliation{Institute for Theoretical Physics and Cosmology, Zhejiang University of Technology, Hangzhou 310023, China}
\affiliation{University of Tashkent for Applied Sciences, Str. Gavhar 1, Tashkent 100149, Uzbekistan}

 \author{Hemwati Nandan
}
\email{hnandan@associates.iucaa.in}
\affiliation{Department of Physics, Hemvati Nandan Bahuguna Garhwal Central University,\\ Srinagar Garhwal, Uttarakhand 246174, India}

\date{\today}
\begin{abstract}

We studied the magnetic Penrose process (MPP) in the Kerr-Bertotti-Robinson (KBR) spacetime, an exact rotating electrovacuum solution describing a black hole (BH) immersed in an intrinsic, uniform electromagnetic field. We analyze the behavior of charged particles in this geometry and find that the spacetime structure itself responds non-monotonically to the background magnetic field $B$. Specifically, both the event horizon and the static limit surface first expand as $B$ increases, reach a maximum size at an intermediate field strength, and then contract toward the extremal limit. Although the ergoregion itself shrinks monotonically with $B$, this structural feature gives rise to a pronounced non-monotonic dependence of the energy extraction efficiency on the magnetic field $B$, i.e., the efficiency initially rises, attains a maximum value, and subsequently falls as the extremal condition is approached. This contrasts sharply with the monotonic trends usually associated with magnetic enhancements in the Kerr geometry. We further explore an astrophysical application of the MPP by estimating the maximum energy of electrons escaping from the ergoregion of the KBR BH. Modeling neutron beta decay occurring near the event horizon, we derive an analytical expression for the energy gained by electrons accelerated by the magnetic field. Applying our results to the supermassive BH at the Galactic center, $\mathrm{SgrA}^*$, we find that electrons can be accelerated up to energies of $\sim 10^{15}\,\mathrm{eV}$ for realistic values of the spin and magnetic field. Although these energies exceed the observed upper range of cosmic-ray electrons, radiative losses such as synchrotron emission and inverse-Compton scattering can efficiently reduce them to the observed $\mathrm{TeV}$ scale. The KBR BH thus provides a clear example of how intrinsically electromagnetic fields can qualitatively affect the BH energetics, introducing an optimal magnetic field strength for rotational energy extraction---a feature with potential implications for high-energy astrophysical environments where magnetic fields are present. Our results show that in electrovacuum spacetimes where the magnetic field is intrinsic to the geometry, the MPP can be significantly modified, revealing a new regime of BH energy extraction.
\end{abstract}

\maketitle

\section{Introduction}
\label{introduction}

The extraction of rotational energy from black holes (BHs) remains one of the most compelling consequences of general relativity, with profound implications for high-energy astrophysics. Since Penrose's seminal proposal that energy could be extracted from a rotating BH via particle disintegration in the ergosphere region \cite{Penrose:1969pc}, numerous mechanisms have been explored to explain the enormous luminosities observed from active galactic nuclei (AGN) \cite{Rees:1984si}, microquasars \cite{Mirabel:1999fy}, and gamma-ray bursts \cite{Blandford1977}. Among these, the Magnetic Penrose Process (MPP) \cite{PhysRevD.29.2712,PhysRevD.30.1625,Wagh85ApJ,1986ApJ...307...38P,Tursunov:2019oiq,Shaymatov24PhRvD.110d4042S} stands out for its remarkable efficiency, potentially exceeding $100\%$ for charged particles in various configuration of magnetic fields \cite{Dadhich:2018gmh,Shaymatov24EPJC...84.1015S,Shaymatov:2022eyz,Khamidov25JCAP...03..053X}.

Recent advancements in gravitational wave astronomy and high-resolution imaging of BH shadows \cite{EventHorizonTelescope:2019dse,EventHorizonTelescope:2022wkp} have revived interest in the fundamental processes that drive BH energetics \cite{Dhang:2024aoa, 2025NatAs.tmp..198F, Punsly:2024dih, Kumar:2025tws,Ghosh:2025vtb, Kim:2024fuy, Rule:2025faz, Prabu:2025nfr}. The MPP, operating through the combined effects of frame-dragging and electromagnetic interactions, offers a robust mechanism for energy extraction that does not require the complex magnetohydrodynamic (MHD) configurations of alternatives like the Blandford-Znajek process \cite{1985JApA....6...85B}. While the Blandford-Znajek mechanism operates through magnetic field lines threading the horizon and is crucial for explaining persistent relativistic jets, the MPP provides a complementary, local mechanism that may dominate during transient events or in environments with specific magnetic field geometries.

The Kerr-Bertotti-Robinson (KBR) spacetime \cite{Podolsky:2025tle} emerges from the fusion of two fundamental solutions in general relativity. The Kerr metric \cite{Kerr:1963ud} describes the vacuum spacetime of a rotating BH, while the Bertotti-Robinson solution \cite{PhysRev.116.1331, 10.1063/1.1703712, Robinson:1959ev} represents a homogeneous electromagnetic universe—a product-type spacetime with geometry AdS\(_2 \times S^2\) that is conformally flat and possesses a constant electromagnetic field. Their combination yields an exact solution to the Einstein-Maxwell equations representing a rotating BH embedded in, and interacting with, a uniform electromagnetic background. This metric was first systematically studied by Carter \cite{PhysRev.174.1559}, who identified it as a member of the broader Pleba\'{n}ski-Demia\'{n}ski class of spacetimes \cite{PLEBANSKI197698,Plebanski:1976zz}.

The importance of the KBR metric lies in its role as a {non-perturbative} description of a magnetized rotating BH. Unlike test-field approaches where the magnetic field is treated as a perturbation on a fixed Kerr background, in the KBR solution the electromagnetic field is an integral part of the geometry, satisfying the field equations self-consistently. This makes it particularly relevant for studying regimes where the magnetic field energy density becomes comparable to the gravitational field energy density near the horizon-conditions that may arise in magnetically dominated accretion flows or in the vicinity of BHs formed from magnetized stellar collapse \cite{Bocquet:1995je}. The metric's non-asymptotically flat structure reflects the presence of a cosmological electromagnetic field that permeates the entire universe, analogous to a uniform magnetic field in cosmological contexts.

Furthermore, due to these intriguing and observationally relevant features, the KBR BH has drawn considerable attention recently, with most investigations focusing on its geometric structure and the observable repercussions of the background electromagnetic field, rather than on energetic particle processes. Considerable work has been devoted to photon motion, geodesics, and BH shadows in the KBR spacetime, demonstrating how a uniform magnetic field modifies null trajectories and optical appearances relative to the Kerr case \cite{Wang:2025vsx,Zeng:2025tji}. Strong-field gravitational lensing and associated frequency shifts induced by magnetization have also been investigated, further highlighting the role of the Bertotti--Robinson electromagnetic background in shaping observational signatures \cite{Vachher:2025jsq}. Complementary analyses of the innermost stable circular orbit and inspiral dynamics reveal that the coupled rotation--electromagnetic structure of the KBR spacetime leads to nontrivial modifications of orbital stability \cite{Wang:2025bjf}.
Beyond purely geometric and optical aspects, several studies have explored dynamical and plasma-related phenomena in the KBR background. In particular, energy release through magnetic reconnection in both circular and plunging plasma flows has been shown to be significantly enhanced by the interplay between rotation and electromagnetic fields \cite{Zeng:2025olq}. The effects of external magnetic fields on spinning test particles orbiting KBR BHs have also been examined, indicating notable deviations from Kerr dynamics \cite{Andersson:2025bhq}. More recently, the KBR metric has appeared in broader theoretical contexts, including horizon-scale tests of gravity based on BH shadows \cite{Liu:2025wwq}, the Kerr/CFT correspondence \cite{Siahaan:2025ngu}, and general classifications of type-D BH spacetimes \cite{Podolsky:2025zlm}. Gravitational-wave imprints of KBR BHs have likewise been analyzed, suggesting that electromagnetic backgrounds can leave measurable signatures in waveform phasing and frequency evolution \cite{Li:2025rtf}.
Despite this growing literature, existing studies predominantly focus on neutral particles, photon dynamics, plasma processes, or indirect energetic effects inferred from orbital or wave behavior. A systematic investigation of energy extraction mechanisms involving {charged} particles in the exact KBR BH spacetime remains notably absent. This gap is particularly significant because the KBR solution represents a fully non-perturbative magnetized rotating BH, where the coupling between rotation and electromagnetic fields is intrinsically strong and cannot be captured by weak-field or test-field approximations. Exploring charged-particle energy extraction in this exact setting is therefore expected to uncover qualitatively new energetic behavior, with direct relevance to high-energy astrophysical processes near magnetized rotating BHs ~\cite{Fender04mnrs,Auchettl17ApJ,IceCube17b}.

The test-field approximation, while computationally convenient and analytically tractable, becomes inadequate when the magnetic field strength approaches or exceeds critical values where backreaction effects become important. For astrophysical BHs, this corresponds to field strengths \(B \sim 10^{8}\) G for stellar-mass BHs or \(B \sim 10^4\) G for supermassive BHs-values \cite{Piotrovich10,Baczko16,Daly:APJ:2019:} that are plausible in magnetar-BH binaries or in the inner regions of certain accretion disk models \cite{2014Natur.510..126Z, Piotrovich:2020ooz}. In such regimes, the magnetic field can drastically influence particle dynamics \cite{Aliev02,Frolov10,Tursunov16,Shaymatov22a,Hussain17,Shaymatov21pdu,2023EPJC...83..323K,Shaymatov22c}. However, in magnetized spacetimes where the magnetic field is intrinsic to the geometry, it not only affects particle dynamics but also modifies the spacetime geometry itself, potentially altering the size and shape of the ergoregion, the location of horizons, and the efficiency of energy extraction mechanisms \cite{Shaymatov21c,Shaymatov:2022eyz,Shaymatov23GRG}.

The KBR spacetime provides an ideal laboratory to study these effects in a controlled, exact setting. Its analytic tractability allows us to isolate the fundamental gravitational-electrodynamic coupling without the complications of accretion physics or time-dependent fields. By studying the MPP in this spacetime, we can answer fundamental questions: does an intrinsically magnetic field always enhance energy extraction, as suggested by test-field calculations? Or does the geometric backreaction introduce limiting or even suppressing effects? How does the ergoregion, the engine room of the Penrose process, evolve when the magnetic field becomes a structural component of spacetime rather than just an environmental factor?

From an observational perspective, understanding energy extraction in magnetized environments is crucial for interpreting data from multi-wavelength observations of BH systems. The non-monotonic behavior we discover-where efficiency peaks at an optimal magnetic field strength—could, in principle, manifest as correlations between jet power and inferred magnetic field strengths in AGN and X-ray binaries \cite{King01ApJ,Peterson:97book}. If the MPP operates alongside or in competition with the Blandford-Znajek mechanism, its distinctive magnetic field dependence might help explain why some systems with very strong magnetic fields show unexpectedly modest jet efficiencies \cite{Nalewajko:2014wqa}.

Moreover, the KBR metric's description of a BH in a uniform electromagnetic universe, while idealized, captures essential physics relevant to BHs in magnetic fields, such as those threading galactic cores or present in binary systems with magnetized companions. The process we describe may be particularly relevant for explaining sudden, high-energy flares from BH systems, which could result from magnetic reconnection events or abrupt changes in accretion flow that temporarily create conditions favorable for the MPP \cite{Karlicky2016}.

In this work, we perform the detailed analysis of the MPP in the exact KBR BH spacetime. Our investigation reveals several novel features that distinguish this system from the test-field Kerr case. Most strikingly, we find that the spacetime structure responds non-monotonically to variations in the background magnetic field \(B\): both the event horizon and static limit surface initially expand with increasing \(B\), reach a maximum size at an intermediate field strength, and then contract toward the extremal limit. This geometric behavior induces a corresponding non-monotonicity in the MPP efficiency, which peaks at an optimal magnetic field strength before declining, a phenomenon absent in standard treatments where efficiency increases monotonically with \(B\).

Our work demonstrates that when magnetic fields are intrinsic to the spacetime geometry, they can qualitatively alter the efficiency landscape of rotational energy extraction. For fixed \(B\), however, the MPP efficiency for negatively charged particles in KBR spacetime consistently surpasses that in comparable Kerr scenarios and increases monotonically with BH spin. These findings suggest that in astrophysical environments where magnetic fields are dynamically important, there may exist optimal conditions for energy extraction that are not captured by perturbative approaches.

The implications of this study extend beyond theoretical interest. With upcoming advancements in multi-messenger astronomy and increasingly precise measurements of BH parameters, understanding how BH spin, magnetic fields, and energy extraction mechanisms work together becomes important to make sense of data from instruments like LISA \cite{2017arXiv170200786A}, the Einstein Telescope \cite{Punturo:2010zz}, and next-generation VLBI arrays \cite{Ayzenberg:2023hfw}. Our results also provide new ways to study the related phenomena, including chaotic scattering of charged particles, ultra-high-energy cosmic ray acceleration, and potential connections to gauge/gravity duality through the AdS\(_2 \times S^2\) structure of the Bertotti-Robinson universe.

The structure of this paper is as follows. We begin in Sec.~\ref{Sec:metric} by introducing the KBR BH metric and examining its key geometric features—specifically the event horizon and ergoregion (with their detailed analytical dependence on the magnetic field provided separately in Appendix~\ref{app:Ana_ergo_B}). Subsequently, Sec.~\ref{sec:geodesic} derives the equations of motion governing charged particles in this KBR BH background. The core analysis is presented in Sec.~\ref{Sec:mpp}, where we formulate the MPP for this spacetime and compute the energy extraction efficiency. We present our principal findings, including the non-monotonic behavior and a comparative analysis, and discuss their astrophysical implications and potential future research directions. Finally, we summarize our conclusions in Sec.~\ref{Sec:conclusion}.

Throughout this paper, we use a spacetime metric signature $(-,+,+,+)$ and adopt geometric units where $G = c = 1$. 

\section{The Kerr-Bertotti-Robinson Spacetime: Horizon and Ergoregion Structure}\label{Sec:metric}

The new exact solution of the Einstein-Maxwell equations describing a Kerr BH immersed in an asymptotically uniform external electromagnetic field-known as the KBR solution was recently presented in Ref.~\cite{Podolsky2025}. The metric describing KBR BH reads as 
\begin{eqnarray}\label{Eq:metric} 
ds^{2} &=& \frac{1}{\omega^{2}} \Big[ -\frac{Q}{\rho^{2}} (dt - a \sin^{2}\theta \, d\varphi)^{2} 
+ \frac{\rho^{2}}{Q} \, dr^{2} + \frac{\rho^{2}}{P} \, d\theta^{2} \nonumber\\ 
&+& \frac{P \sin^{2}\theta}{\rho^{2}} (a \, dt - (r^{2}+a^{2}) \, d\varphi)^{2} \Big],
\end{eqnarray}
with
\begin{eqnarray}\label{notations}
\rho^{2} &=& r^{2} + a^{2}\cos^{2}\theta, \quad \nonumber\\
P &=& 1 + B^{2}\!\left( m^{2} \frac{I_{2}}{I_{1}^{2}} - a^{2} \right)\cos^{2}\theta, \quad \nonumber\\
Q &=& (1 + B^{2}r^{2})\,\Delta, \quad \nonumber\\
\omega^{2} &=& (1 + B^{2}r^{2}) - B^{2}\Delta\cos^{2}\theta, \quad \nonumber\\
\Delta &=& \left( 1 - B^{2}m^{2}\frac{I_{2}}{I_{1}^{2}} \right)r^{2} - 2mr \frac{I_{2}}{I_{1}} + a^{2}, \quad \nonumber\\
I_{1} &=& 1 - \tfrac{1}{2}B^{2}a^{2}, \quad 
I_{2} = 1 - B^{2}a^{2}\, ,
\end{eqnarray}
where, $m$ is the mass of the BH, $a$ is the spin parameter, and $B$ is the magnetic field strength. When  $B = 0$, the metric reduces to the Kerr metric, when   $m = 0$, it reduces to the Bertotti-Robinson metric, and when $a = 0$, it reduces to the Schwarzschild-Bertotti-Robinson metric.

The structure of horizons in the KBR BH spacetime is
determined by the roots of the generalized radial function $\Delta$, given by
\begin{equation}
\Delta = 
\left( 1 - B^{2}m^{2}\frac{I_{2}}{I_{1}^{\,2}} \right) r^{2}
- 2mr\,\frac{I_{2}}{I_{1}} + a^{2}\, ,
\end{equation}
where $I_{1} = 1 - \tfrac{1}{2}B^{2}a^{2}$ and $I_{2} = 1 - B^{2}a^{2}$.
Since $\Delta$ is a quadratic function of $r$, the existence and nature of
horizons are controlled by the discriminant of this quadratic,
\begin{equation}
D(a,B) = 
\left[-2m\,\frac{I_{2}}{I_{1}}\right]^{2}
- 4\left(1 - B^{2}m^{2}\frac{I_{2}}{I_{1}^{\,2}}\right)a^{2}\, .
\label{eq:discriminant}
\end{equation}

\begin{figure}[ht!]
\centering
\includegraphics[width=0.5\textwidth]{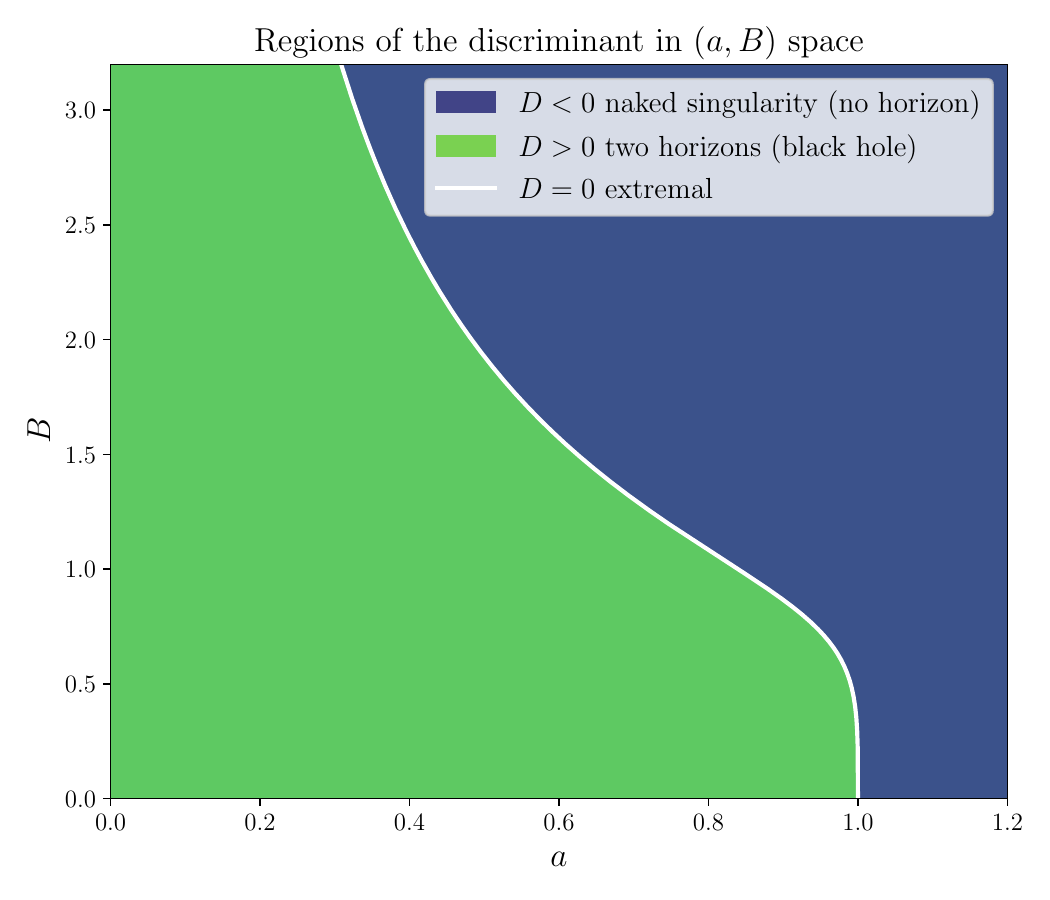}
\caption{Parameter space of the KBR BH in the
$(a,B)$ plane.  
The shaded regions indicate the sign of the discriminant $D$ of the
radial function $\Delta$:  
the region with $D>0$ corresponds to BHs with two horizons,
$D=0$ marks the extremal curve (shown as a white solid line)
and $D<0$ corresponds to configurations without horizons.
}
\label{fig:DeltaRegion}
\end{figure}

The sign of the discriminant governs the causal structure:
\begin{enumerate}[label=(\roman*)]
    \item $D>0$ : two distinct real roots $\Rightarrow$ inner and outer horizons exist,
    \item $D=0$ : double root $\Rightarrow$ extremal KBR BH,
    \item $D<0$ : no real roots $\Rightarrow$ horizonless configuration.
\end{enumerate}
Because $D$ depends only on the spin $a$ and the external field strength $B$,
the $(a,B)$ parameter space can be divided into horizon and no-horizon
regions, as seen in Fig.~\ref{fig:DeltaRegion}.  In the Kerr limit ($B=0$), one recovers the well-known condition
$a\le m$ for the existence of horizons.  For a nonzero $B$, however, the allowed
parameter region is significantly deformed, reflecting the interplay between
rotation and the external magnetic field.

\begin{figure*}
\centering
\includegraphics[scale=0.72]{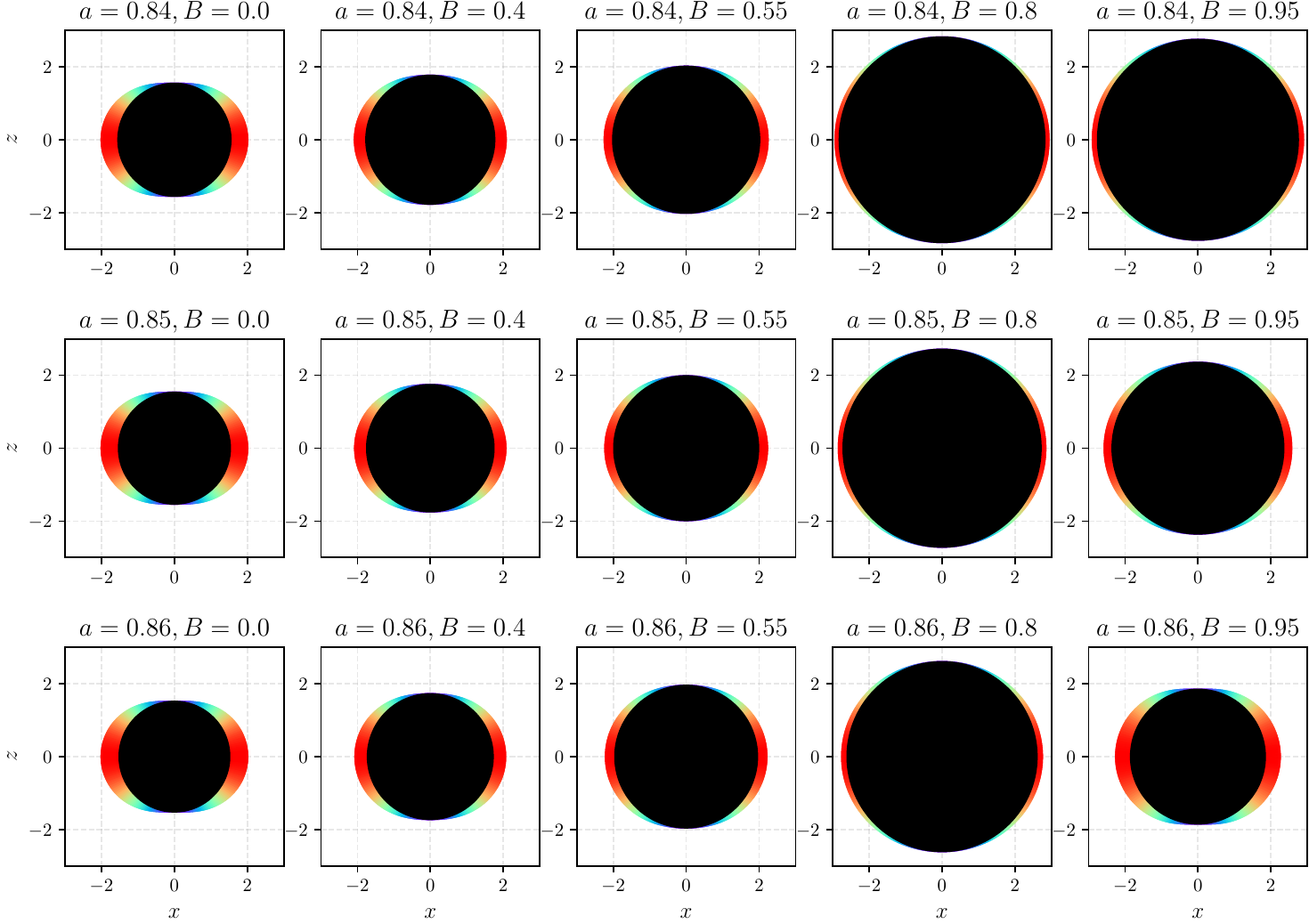}
\caption{
Event horizon and static limit surface (ergosphere) of the KBR BH for 
different values of the spin parameter $a$ and magnetic field parameter $B$ are shown in the $x-z$ plane. 
Along each {row}, the parameter $B$ increases while the 
spin parameter $a$ is held fixed; along each {column}, the parameter $a$ 
increases for a fixed value of $B$. 
The black region denotes the event horizon, while the coloured region between the 
horizon and the static limit marks the ergoregion. Red colour indicates the 
maximum ergoregion thickness near the equator, gradually shifting to blue as it 
decreases toward the poles.
}
\label{fig:Ergosphere}
\end{figure*}

It is worth emphasizing that the external magnetic field modifies the
effective mass and spin contributions inside $\Delta$, so the boundary between
BH and no-BH states is no longer linear as in the Kerr case. The deformation of the extremality curve revealed in
Fig.~\ref{fig:DeltaRegion} originates from the $B$--dependent rescalings of the coefficients of the quadratic term in $r$.  
In particular, the quantities $I_{1}$ and $I_{2}$ play the role of
magnetically induced renormalizations of the spin term and the mass term,
respectively, leading to nontrivial coupled constraints on $(a,B)$.

In addition to the horizons, the KBR spacetime possesses a static limit surface, defined as the outer boundary of the region where no static observer can remain at rest with respect to infinity (i.e. the ergoregion). The static limit is determined by the condition $g_{tt}=0$, which gives the equation
\begin{align}\label{eq:static_limit}
g_{tt} &= -\frac{Q}{\rho^{2}} + \frac{P \, a^{2}\sin^{2}\theta}{\rho^{2}} = 0, \nonumber\\ 
&\Rightarrow(1 + B^{2} r^{2}) \Delta - a^{2} P \sin^{2}\theta = 0\, ,
\end{align}
where $\Delta$ and $P$ are defined as in Eq.~\eqref{notations}. Solving Eq.~\eqref{eq:static_limit} for $r$ gives the radius of the static limit surface $r_{\rm SLS}(\theta)$ as a function of the polar angle $\theta$. At poles $(\theta=0,\pi)$, the static limit coincides with the event horizon $r_{\rm SLS} = r_h$, while at the equatorial plane $(\theta=\pi/2)$, the static limit surface reaches its maximum radius, creating the characteristic ergosphere around the BH.
\begin{figure*}[htb!]
\centering
    \includegraphics[width=1.0\textwidth]{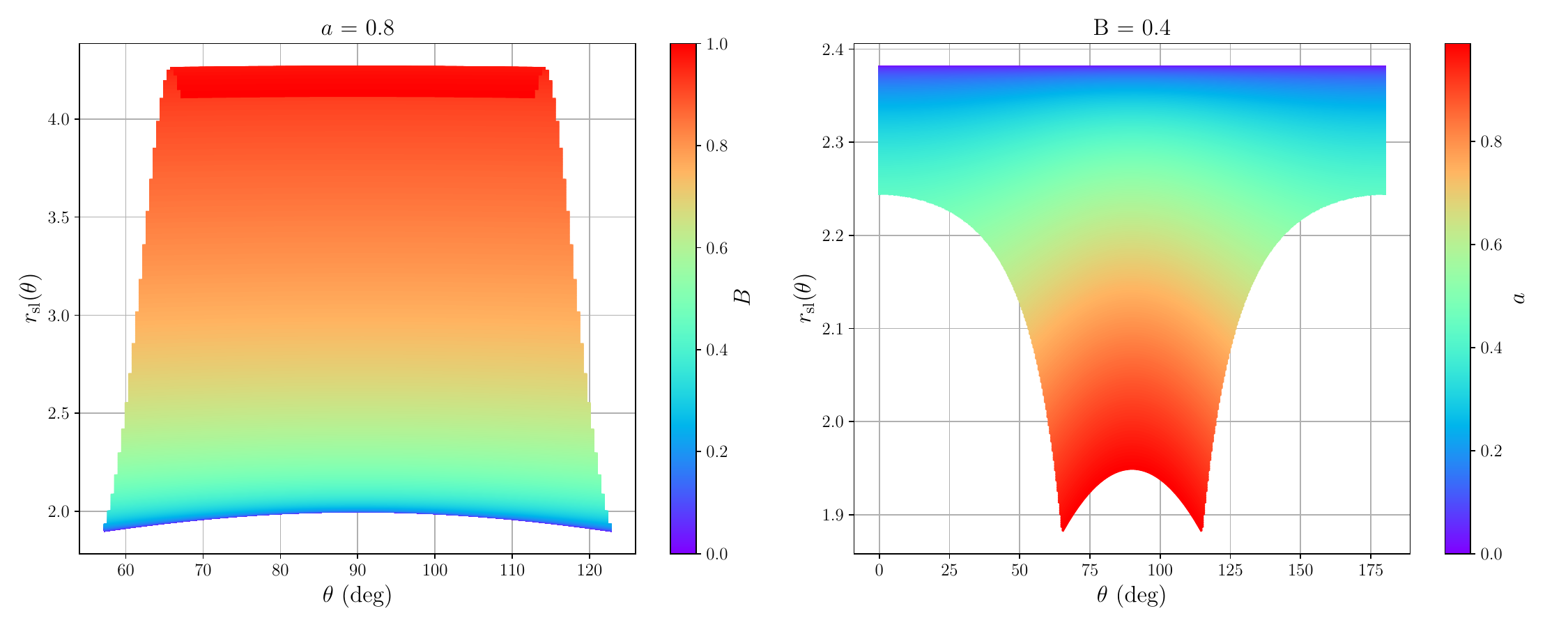}\\
    \includegraphics[width=0.49\linewidth]{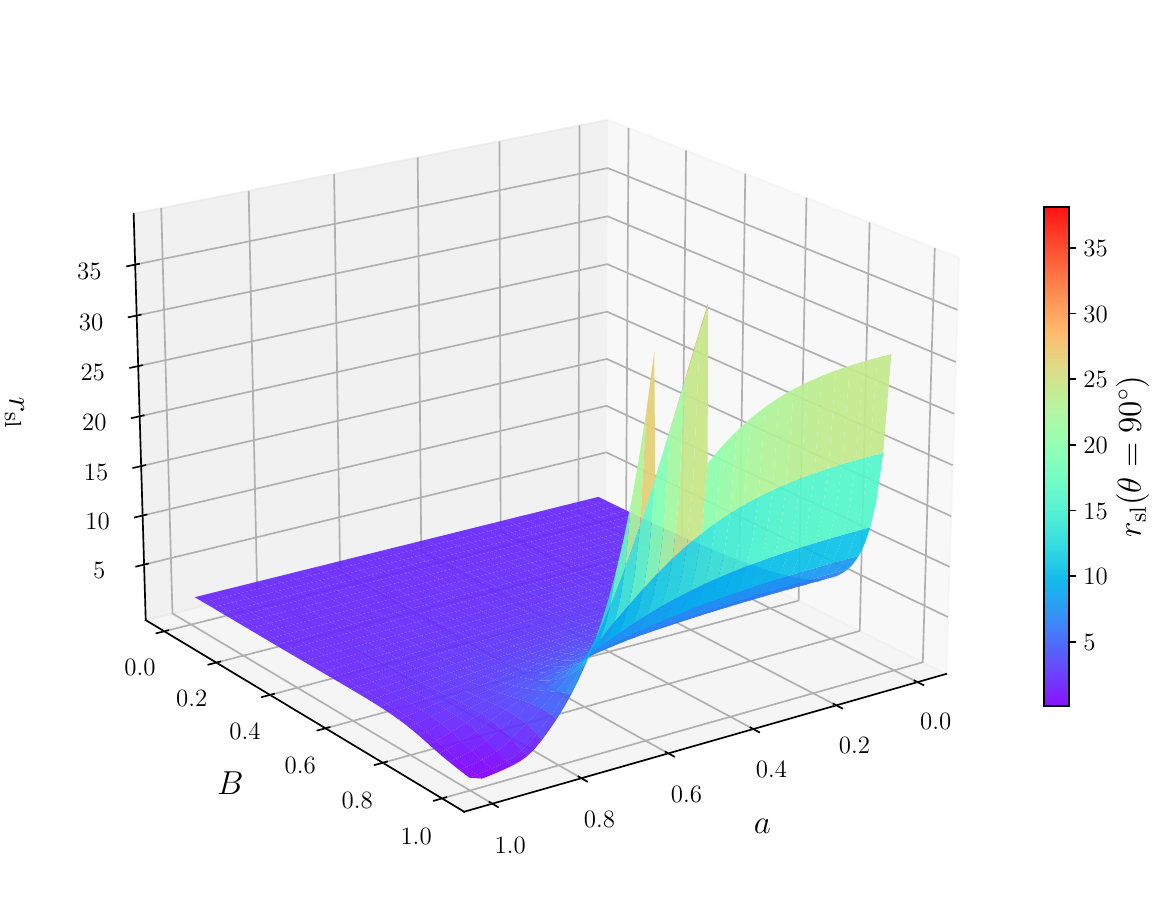}
    \includegraphics[width=0.5\linewidth]{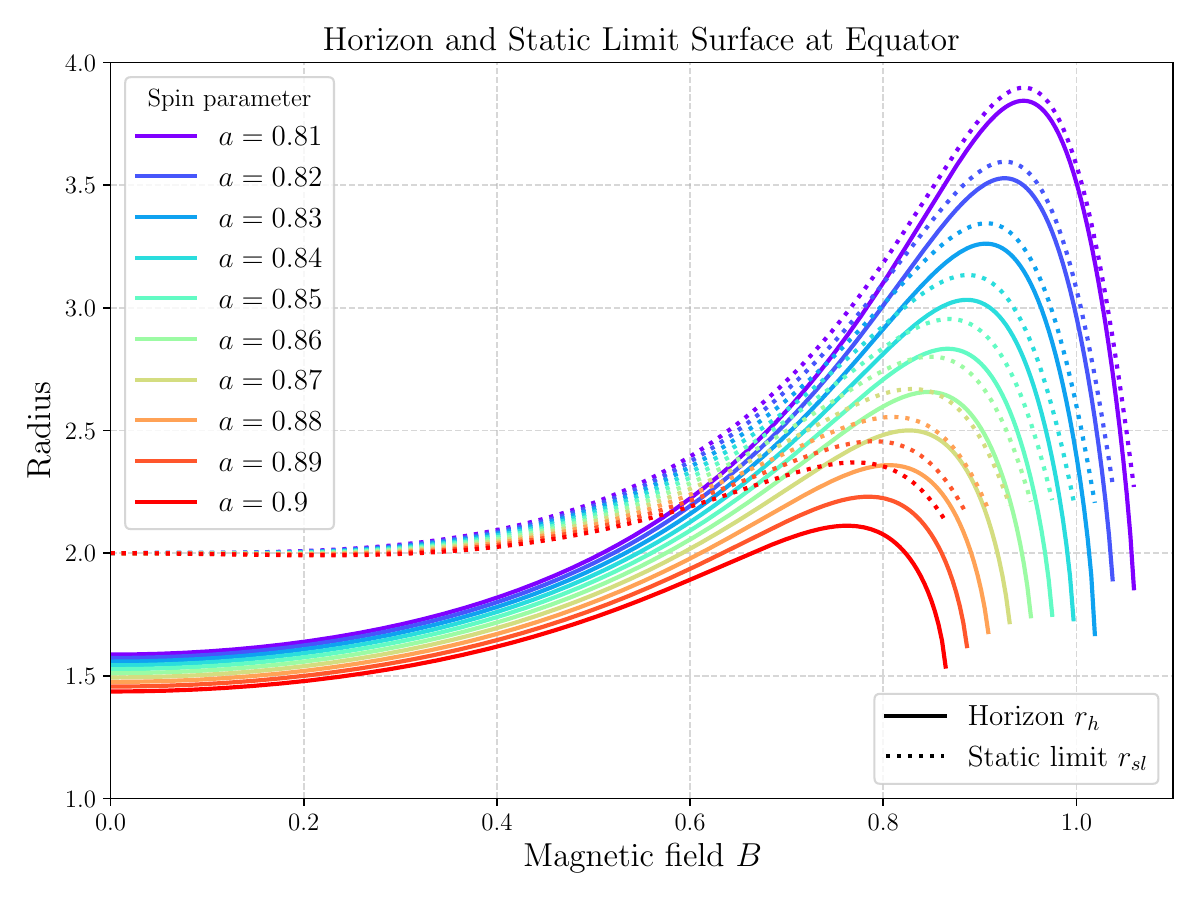}
\caption{
Top left panel: Angular dependence of the static limit surface 
$r_{\rm sl}(\theta)$ for different values of the $B$ parameter at fixed spin $a=0.8$. Top right panel: shows the angular dependence of the static limit surface 
$r_{\rm sl}(\theta)$ for different values of the spin parameter $a$ at fixed spin $B$.
Bottom left panel: Equatorial static limit radius 
$r_{\rm sl}(\theta=\pi/2)$ across the $(a,B)$ parameter space. 
{Bottom-right panel:} Equatorial static limit radius
$r_{\rm sl}(\theta=\pi/2)$ and horizon radius $r_{\rm h}$ as a function of parameter $B$ for different values of spin parameter $a$.
}
\label{fig:sls_combined}
\end{figure*}

The ergoregion lies between the event horizon and the static limit surface. Within this region, particles cannot remain stationary with respect to asymptotic infinity and must co-rotate with the BH. Here, the shape and size of the ergosphere depend on both the spin $a$ and the external magnetic field $B$. Increasing $a$ enlarges the equatorial extension, while increasing $B$ modifies the shape through magnetically induced factors $I_1$ and $I_2$ in $\Delta$ and $P$, leading to nontrivial deformations compared to the Kerr BH case. In the equatorial plane $(\theta=\pi/2)$, the static limit radius can be explicitly expressed as follows: 
\begin{equation}
r_{\rm sl}^{\rm eq} = \frac{ m I_{2}/I_{1} + \sqrt{ (m I_{2}/I_{1})^{2} + a^{2} (1 - B^{2} m^{2} I_{2}/I_{1}^{2}) }}{1 - B^{2} m^{2} I_{2}/I_{1}^{2}}\, .
\end{equation}
This expression reduces to the standard Kerr static limit $r_{\rm sl}^{\rm eq} = m + \sqrt{m^{2}-a^{2}}$ when $B=0$.

As illustrated in Fig.~\ref{fig:Ergosphere}, the external magnetic field has a non-monotonic influence on the size of the ergoregion. For fixed spin $a$, the equatorial thickness of the ergosphere initially grows as $B$ increases, reflecting the enhancement of frame dragging due to the magnetically modified coefficients $I_{1}$ and $I_{2}$. However, beyond a certain field strength, the deformation reverses and the ergoregion begins to shrink again. This nontrivial behavior arises from the $B$--dependent competition between the effective mass and rotation terms appearing in $\Delta$ and $P$, leading to a maximum ergoregion size at intermediate magnetic field values.

The behavior of the static limit surface is further illustrated in 
Fig.~\ref{fig:sls_combined}. In the top left panel, we show the full
angular dependence $r_{\rm SLS}(\theta)$ for different values of the external magnetic field for keeping $a=0.8$ fixed. The equatorial radius increases more rapidly than the polar radius, indicating that the ergoregion becomes increasingly oblate as $B$ grows. In the top right panel, we show the angular dependence $r_{\rm SLS}(\theta)$ for keeping the external magnetic field $B=0.4$ fixed and for varying the spin parameter $a$. This effect is quantified in the bottom right panel of Fig.~\ref{fig:sls_combined}, which maps the static equatorial limit radius $r_{\rm SLS}^{\rm eq}$ over the parameter space $(a,B)$. The equatorial ergoregion exhibits a non-monotonic response: it expands with increasing $B$, reaches a maximum, and then decreases for larger field strengths (for an analytical treatment, see \ref{app:Ana_ergo_B}). This arises from the competition between the magnetically rescaled mass and spin contributions entering $\Delta$ and $P$, leading to a maximal ergoregion size at intermediate $B$, as shown in the bottom right panel.  

\section{DYNAMICS of charged particle AROUND KBR BH SPACETIME}\label{sec:geodesic}

To investigate how a charged particle behaves near the KBR BH, we adopt the Hamilton formalism. In this approach, the Hamiltonian is written as~\cite{Misner73}
\begin{eqnarray}
H=\frac{1}{2}g^{\mu\nu} \left(P_{\mu}-qA_{\mu}\right)\left(P_{\nu}-qA_{\nu}\right)\, ,
\end{eqnarray}
where $P_{\mu}$ corresponds to the particle’s canonical momentum, while $A_{\mu}$ is the electromagnetic four-vector potential. 
The real part of the electromagnetic four-potential can be written as follows~\cite{Podolsky2025}:  
\begin{align}
A_t &= \frac{a (r\, \omega_{,r} +\omega_{,\theta} \cot\theta)}{ B \left(r^2+a^2 \cos ^2\theta \right)}\, , \\
A_{\phi} &=-\frac{r \left(\omega_{,r} \left(a^2+r^2\right)-\omega\, r+r\right)}{ B \left(r^2+a^2 \cos ^2\theta \right)}\nonumber\\
&-\frac{a^2 \cos^2\theta (1-\omega) +a^2 \omega_{,\theta} \sin\theta\cos\theta }{ B \left(r^2+a^2 \cos ^2\theta \right)}\, ,
\end{align}
where $\omega$ is given in Eq.~\eqref{notations}. The four-momentum and canonical momentum of a charged particle are related through $p^{\mu}=g^{\mu\nu}\left(P_{\nu}-qA_{\nu}\right)\, .$
Its motion is governed by the Hamilton equations,
\begin{eqnarray} 
\label{Eq:eqh1}
  \frac{dx^\alpha}{d\lambda} = \frac{\partial H}{\partial P_\alpha}\,   \mbox{~~and~~}
  \frac{dP_\alpha}{d\lambda} = - \frac{\partial H}{\partial x^\alpha}\, , 
\end{eqnarray}
where $\lambda = \tau / \tilde{m}$ denotes the affine parameter corresponding to the particle’s proper time $(\tau)$ along a timelike trajectory. 

\begin{figure*}[ht]
\includegraphics[scale=0.62]{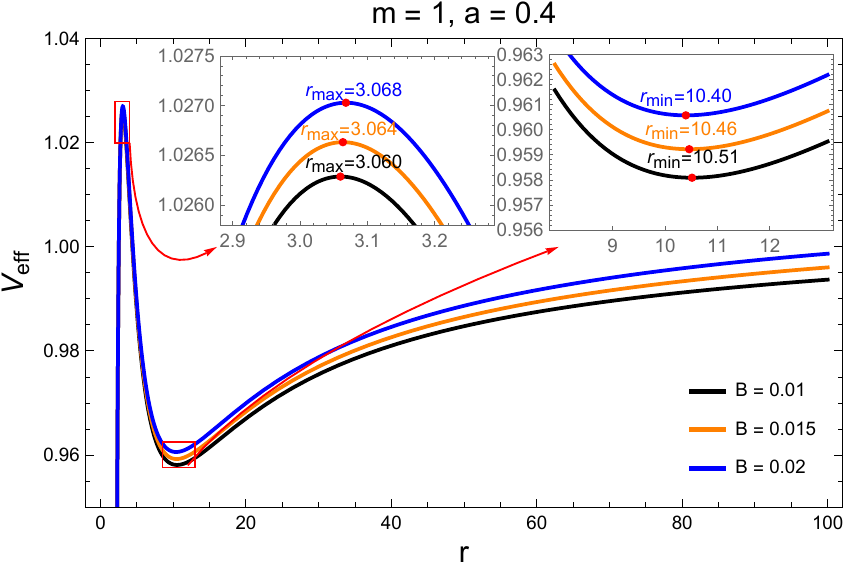}
\includegraphics[scale=0.55]{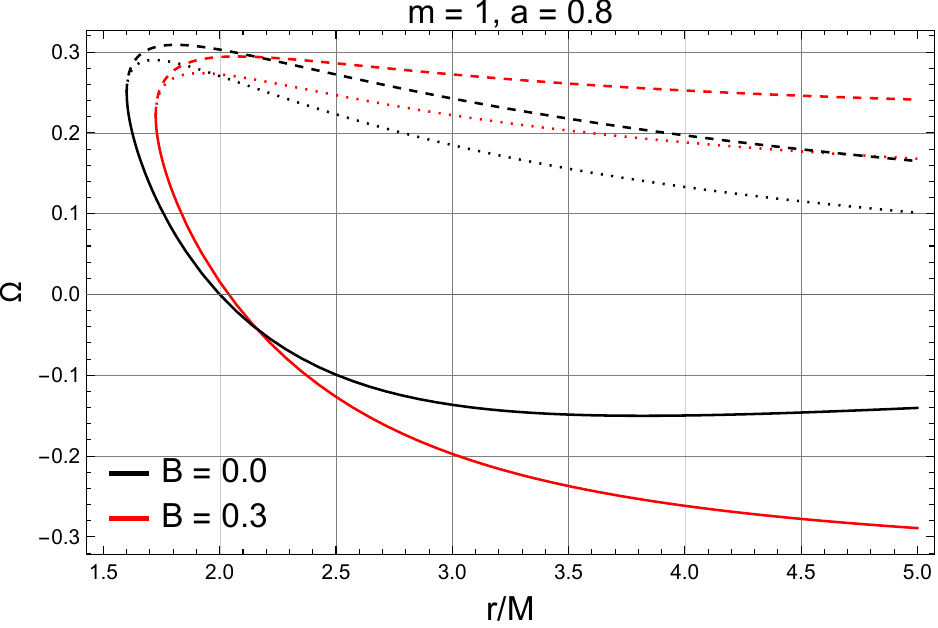}
	\caption{\label{fig:effpot} 
 {Left panel:} Radial profile of the effective potential $V_{\text{eff}}$ for different values of the magnetic field $B$. 
{Right panel:} Radial profiles of $\Omega_{+}$, $\Omega_{-}$, and $\Omega$ for different values of the magnetic field $B$. 
Solid lines correspond to $\Omega_{-}$, dashed lines to $\Omega_{+}$, and dotted lines to $\Omega$.
}
\end{figure*}

From Eq.~\eqref{Eq:eqh1} and the normalization $g_{\mu\nu}p^{\mu}p^{\nu} = -\tilde{m}^2$, one can define the effective potential for timelike radial motion in the equatorial plane ($\theta = \pi/2$) as
\begin{align}\label{Eq:Veff}
V_{\mathrm{eff}} ={}&
- \frac{a B q/\tilde{m}}{m \sqrt{1 + B^2 r^2}}
- \frac{a (-a^2 - r^2 + k)\Lambda}
{(a^2 + r^2)^2 - a^2 k}\nonumber \\[6pt]
&\hspace{-3em}
+ \Bigg\{
\left[
- \frac{a^2 - k}{r^2 (1 + B^2 r^2)}
+ \frac{a^2 (-a^2 - r^2 + k)^2}
{r^2 (1 + B^2 r^2)\bigl((a^2 + r^2)^2 - a^2 k\bigr)}
\right] \nonumber\\[6pt]
&
\times
\left[
1 + \frac{r^2 (1 + B^2 r^2)\Lambda^2}
{(a^2 + r^2)^2 - a^2 k}
\right]
\Bigg\}^{1/2},
\end{align}
where
\begin{eqnarray}
k &=& (1 + B^2 r^2)\left[
a^2
- \frac{2(1 - a^2 B^2) m r}{1 - \frac{a^2 B^2}{2}} 
\right] \nonumber\\[6pt]
&+& \left(
1 - \frac{B^2 (1 - a^2 B^2) m^2}{\left(1 - \frac{a^2 B^2}{2}\right)^2}
\right) r^2,
\end{eqnarray}
\begin{equation}
    \Lambda =
\mathcal{L}
+ \frac{q/\tilde{m}}{B m r}
\left[
r
+ \frac{B^2 r (a^2 + r^2)- r(1 + B^2 r^2)}{\sqrt{1 + B^2 r^2}}
\right]\, .
\end{equation}
where $\mathcal{L} = L/\tilde{m}$ is the particle’s specific angular momentum. The radial motion of a particle can be easily analyzed using the effective potential $V_{\text{eff}}$. The left panel of Fig.~\ref{fig:effpot} shows the radial behavior of the effective potential $V_{\text{eff}}$ for a particle 
moving in the vicinity of the KBR BH. As shown in the figure, increasing the magnetic field causes the minimum and maximum of the effective potential $V_{\text{eff}}$, associated with stable and unstable orbits, to shift toward smaller and larger radii, respectively. This indicates that, for a charged particle, an increase in $B$ produces an additional attractive force directed toward the BH.

We now analyze the specific energy and angular momentum of charged particles on the innermost stable circular orbit (ISCO), $\mathcal{E}_{\rm ISCO}$ and $\mathcal{L}_{\rm ISCO}$. These quantities follow from the simultaneous solution of the following equations \cite{Dadhich22a,Dadhich22IJMPD}
\begin{equation}
    V_{\rm eff}(r_{\rm ISCO}) = \mathcal{E}_{\rm ISCO}^2, 
\quad 
\frac{\partial V_{\rm eff}(r)}{\partial r} = 
\frac{\partial^2 V_{\rm eff}(r)}{\partial r^2} = 0.
\end{equation}
\renewcommand{\arraystretch}{1.0}
\begin{table}[h!]
\centering
\resizebox{0.4\textwidth}{!}{
\begin{tabular}{|c|c|c|c|c|}
\hline
ISCO & \diagbox{$B$}{$a$} & 0 & 0.4 & 0.8 \\ \hline

\multirow{3}{*}{$r_{\mathrm{ISCO}}$}
  & 0.000      & 6.00000       & 4.61434  & 2.90664 \\ 
  & 0.010  & 5.98948  & 4.61103  & 2.90629 \\ 
  & 0.015  & 5.95934  & 4.60130 & 2.90522 \\ \hline

\multirow{3}{*}{$\mathcal{L}_{\mathrm{ISCO}}$}
  & 0.000      & 3.46410   & 3.03407 & 2.38044 \\ 
  & 0.010  & 3.49682  & 3.05509 & 2.38914 \\ 
  & 0.015  & 3.53509   & 3.07862 & 2.39847 \\ \hline

\multirow{3}{*}{$\mathcal{E}_{\mathrm{ISCO}}$}
  & 0.000      & 0.94281 & 0.92494  & 0.87786 \\ 
  & 0.010  & 0.95136 & 0.93032 & 0.87962 \\ 
  & 0.015  & 0.96071 & 0.93614 & 0.88152 \\ \hline

\end{tabular}
}
\caption{ISCO radius, specific angular momentum, and specific energy for different values of magnetic field $B$ and spin parameter $a$. The mass of the KBR BH is fixed at $m=1$, and the particle charge is set to $q=1$.}\label{tab:isco}
\end{table}
\renewcommand{\arraystretch}{1.0}
\begin{table}[h!]
\centering
\resizebox{0.4\textwidth}{!}{
\begin{tabular}{|c|c|c|c|c|}
\hline
ISCO & \diagbox{$B$}{$q$} & -1 & 0 & 1 \\ \hline

\multirow{3}{*}{$r_{\mathrm{ISCO}}$}
  & 0.000      & 3.39313       & 3.39313  & 3.39313 \\ 
  & 0.010  & 3.39235  & 3.39321  & 3.39235 \\ 
  & 0.015  & 3.39006  & 3.39347 & 3.39001 \\ \hline

\multirow{3}{*}{$\mathcal{L}_{\mathrm{ISCO}}$}
  & 0.000      & 2.58650   & 2.58650 & 2.58650 \\ 
  & 0.010  & 2.57554  & 2.58667 & 2.59843 \\ 
  & 0.015  & 2.56550   & 2.58716 & 2.61137 \\ \hline

\multirow{3}{*}{$\mathcal{E}_{\mathrm{ISCO}}$}
  & 0.000      & 0.89639 & 0.89639  & 0.89639 \\ 
  & 0.010  & 0.89386 & 0.89643 & 0.89914 \\ 
  & 0.015  & 0.89152 & 0.89653 & 0.90210 \\ \hline

\end{tabular}
}
\caption{ISCO radius, specific angular momentum, and specific energy for different values of magnetic field $B$ and the particle charge $q$. The mass of the KBR BH is fixed at $m=1$, and spin parameter is set to $a=0.7$.}\label{tab:isco2}
\end{table}

The values of the ISCO parameters for different magnetic fields $B$ and spin parameters $a$ are listed in Table~\ref{tab:isco}. From Table~\ref{tab:isco}, one can see that as the spin parameter $a$ increases, the ISCO parameters decrease due to the enhancement of the centrifugal effect. This behavior is a well-known consequence of BH rotation. On the other hand, an increase in the magnetic field leads to an increase in the ISCO radius, indicating the presence of an additional attractive force directed toward the BH, as shown in Tables~\ref{tab:isco} and \ref{tab:isco2}.
This result is consistent with our conclusions from the effective potential analysis.

We assume that a particle moving in the equatorial plane ($\theta = \pi/2$) follows a circular orbit of constant radius $r$. In this case, the radial and polar components of the velocity vanish, $dr/dt = d\theta/dt = 0$, and the angular velocity is defined as $\Omega = d\phi/dt = u^\phi/u^t$. Using the normalization condition for a photon, we obtain the following expression~\cite{1986ApJ...307...38P,Wagh:1985vuj}
\begin{align}
\Omega_{\pm}
= \frac{-g_{t\phi} \pm \sqrt{(g_{t\phi})^{2} - g_{tt} g_{\phi\phi}}}{g_{\phi\phi}}\, ,
\end{align}
where the $\pm$ signs correspond to the corotating ($+$) and counterrotating ($-$) photon orbits. Using the normalization condition for a timelike particle, we can also obtain the following expression~\cite{1986ApJ...307...38P,Shaymatov:2022eyz,Nozawa05}
\begin{eqnarray}\label{29}
\Omega=\frac{-g_{t\phi}\left(p_t^2+g_{tt}\right)+\sqrt{\left(p_t^2+g_{tt}\right)\left(g_{t\phi}^2-g_{tt}g_{\phi\phi}\right)p_t^2}}{g_{\phi\phi}p_t^2+g_{t\phi}^2}\, .\nonumber\\
\end{eqnarray}
where we have denoted $p_t=-\left(\mathcal{E}+qA_{t}/\tilde{m}\right)$. The angular velocity of a timelike particle is bounded by $\Omega_{-} < \Omega < \Omega_{+}$. The angular velocities $\Omega$ and $\Omega_{+}$ are always positive, while $\Omega_{-}$ can take both positive and negative values, as shown in the right panel of Fig.~\ref{fig:effpot}.

\section{The Magnetic Penrose Process: Theory of Energy Extraction and Its Astrophysical Relevance}\label{Sec:mpp}

The idea that energy could be extracted from a BH was first demonstrated by Roger Penrose \cite{Penrose:1969pc}. Further research on how magnetic fields affect the energy extraction mechanism has contributed to explaining numerous high-energy astrophysical events~\cite{Wagh85ApJ, 1985JApA....6...85B,Comisso21MR,Shaymatov23MRP}.

In Penrose’s original model \cite{Penrose:1969pc,Abdujabbarov11}, a massive particle with energy $E_1$ enters the ergoregion and splits into two fragments with energies $E_2$ and $E_3$. One of these fragments, say the one with energy $E_2$, falls into the BH, while the other fragment, with energy $E_3$, escapes to infinity. The efficiency of this energy extraction process is given by the following expression
\begin{eqnarray} \label{Eq:eta}
\eta=\frac{E_3-E_1}{E_1}\, .
\end{eqnarray}
\begin{figure*}
\includegraphics[scale=0.6]{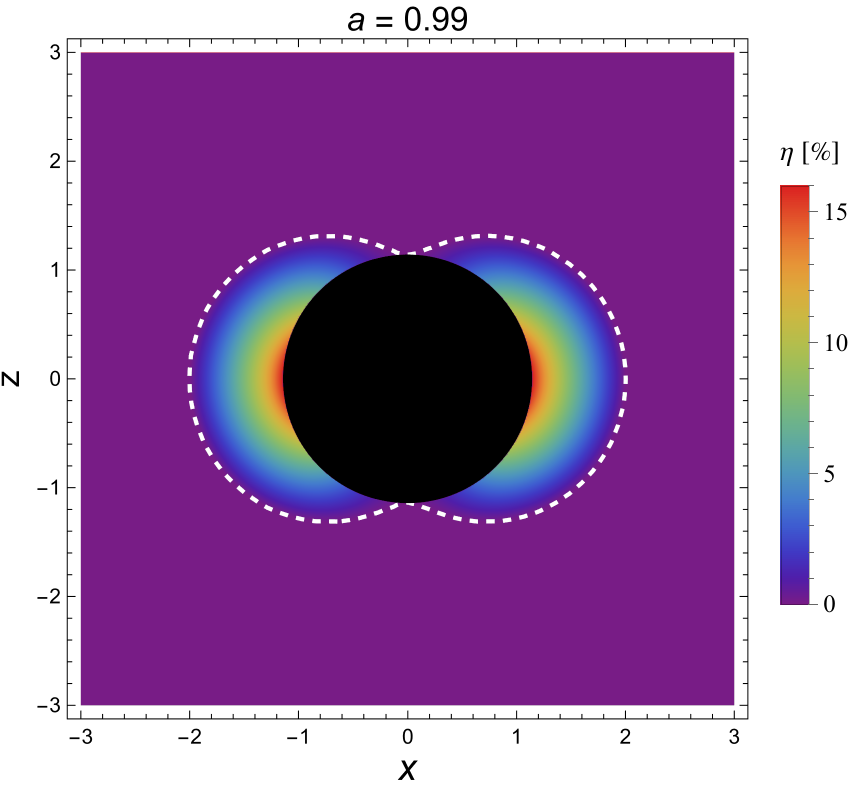}
\includegraphics[scale=0.6]{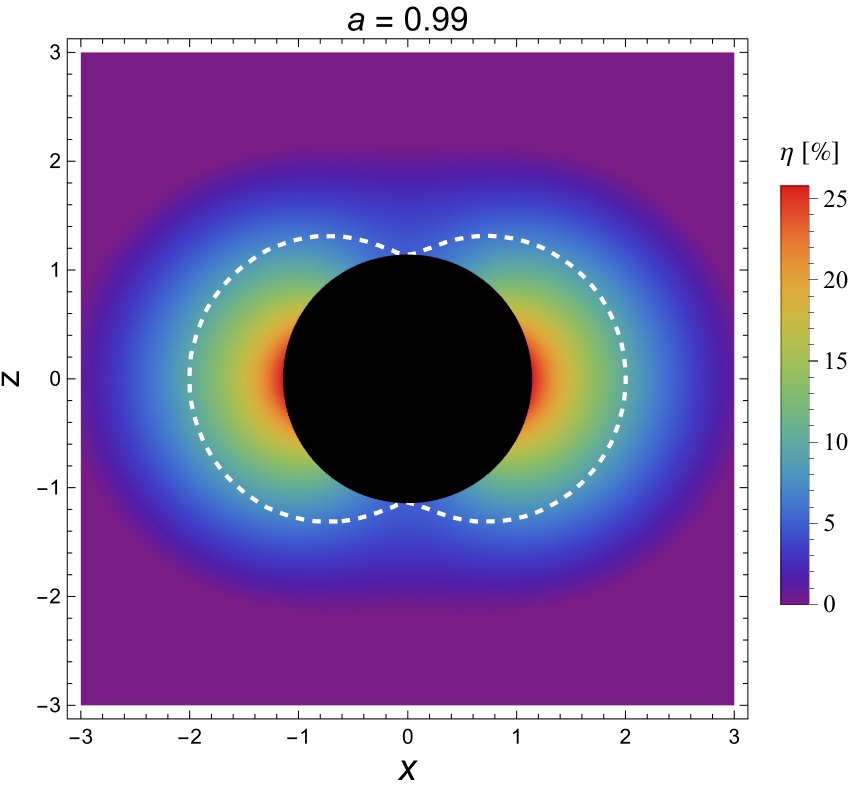}
\caption{\label{fig:en_effKBR} \label{fig:en_eff_dist} Efficiency distribution of energy extraction via the MPP for the KBR BH. 
The left panel shows the case $q_3/m_1 = 0$, while the right panel corresponds to $q_3/m_1 = -2$. The magnetic field parameter is fixed at $B = 0.05$. }
\end{figure*}
In the MPP, a particle with energy and charge $(E_1, q_1)$ decays within the ergoregion into two fragments denoted by $(E_2, q_2)$ and $(E_3, q_3)$. As in the geometric Penrose process, one fragment—say the particle with $(E_2, q_2)$ - falls into the BH, while the other fragment, $(E_3, q_3)$, escapes to infinity. In this case, however, the efficiency of energy extraction differs from that of the purely geometric Penrose process, due to the additional electromagnetic interactions between the charged particles and the magnetic field. For simplicity, we assume that initial neutral particle ($q_1=0$) decays into two parts within the ergoregion. We can write the following conservation laws for this process~\cite{Shaymatov:2022eyz}:
\begin{eqnarray}
    E_1&=&E_{2}+E_{3}\, ,\\
    L_1&=&L_{2}+L_{3}\, ,\\
\label{Eq:con_law}
    m_1u_1^{\mu}&=& m_2u_2^{\mu}+m_3u_3^{\mu}\, ,\\
    q_1 &=& q_2+q_3 = 0\, ,
\end{eqnarray}
where $u_i^{\mu}$ ($i = 1, 2, 3$) denotes the four-velocity of the $i$-th particle. Using the relation $u_i^{\phi}=\Omega_i\, u_i^{t} = -\Omega_i \Lambda_i/\Gamma_i$ and substituting it into Eq.~(\ref{Eq:con_law}), we obtain
\begin{eqnarray}
\Omega_1 m_1 \Lambda_{1} \Gamma_2 \Gamma_3
= \Omega_2 m_2 \Lambda_{2} \Gamma_3 \Gamma_1
+ \Omega_3 m_3 \Lambda_{3} \Gamma_2 \Gamma_1\, ,
\end{eqnarray}
where the quantities $\Lambda_i$ and $\Gamma_i$ are defined as $\Lambda_i = \mathcal{E}_i + q_i A_{t}/m_i$ and $\Gamma_i = g_{tt} + g_{t\phi} \Omega_i$.  
From the above expression, we obtain
\begin{eqnarray}
\frac{E_3+q_3A_{t}}{E_1+q_1A_{t}}=\left(\frac{\Omega_1\Gamma_2-\Omega_2\Gamma_1}{\Omega_3\Gamma_2-\Omega_2\Gamma_3}\right)\frac{\Gamma_3}{\Gamma_1}\, ,
\end{eqnarray}
Now, the energy of the escaping particle can be written as
\begin{eqnarray}\label{Eq:E3E1}
E_3 = \left(\frac{\Omega_1 - \Omega_2}{\Omega_3 - \Omega_2}\right)
\frac{\Gamma_3}{\Gamma_1} \left(E_1 + q_1 A_{t}\right) - q_3 A_{t}\, .
\end{eqnarray}
Using Eqs.~\eqref{Eq:E3E1} and \eqref{Eq:eta}, and assuming $q_1 = 0$, we can express the efficiency as
\begin{eqnarray}
\eta = \left(\frac{\Omega_1 - \Omega_2}{\Omega_3 - \Omega_2}\right)
\frac{\Gamma_3}{\Gamma_1}  - \frac{q_3 A_t}{E_1}- 1\, .
\end{eqnarray}
To maximize the efficiency, the conditions $\Omega_1 = \Omega$, $\Omega_2 = \Omega_{-}$, and $\Omega_3 = \Omega_{+}$ must be satisfied. Consequently, the efficiency of energy extraction from the KBR BH via the MPP can be written as
\begin{eqnarray} \label{Eq:EnerEff}
\eta = \left(\frac{\Omega - \Omega_{-}}{\Omega_{+} - \Omega_{-}}\right)
\left(\frac{g_{tt} + \Omega_{+} g_{t\phi}}{g_{tt} + \Omega\, g_{t\phi}}\right)
- 1 - \frac{q_3 A_t}{E_1}\, .
\end{eqnarray}
If the decay occurs in the equatorial plane ($\theta = \pi/2$) and at the horizon $r_h$, the efficiency reaches its maximum possible value
\begin{widetext}
\begin{equation} \label{En:effmax}
    \eta_{max} = \frac{1}{2}\left(\sqrt{\frac{4 m \left(a^2 B^2-1\right)}{r_h \left(a^2 B^2-2\right)}-\frac{4 B^2 m^2 \left(a^2 B^2-1\right)}{\left(a^2 B^2-2\right)^2}-\frac{a^2 B^2}{B^2 r^2_h+1}}-1\right) - \frac{q_3}{m_1}\frac{aB}{\sqrt{1+B^2r_h^2}}\, .
\end{equation}
\end{widetext}
\begin{figure*}
\includegraphics[scale=0.5]{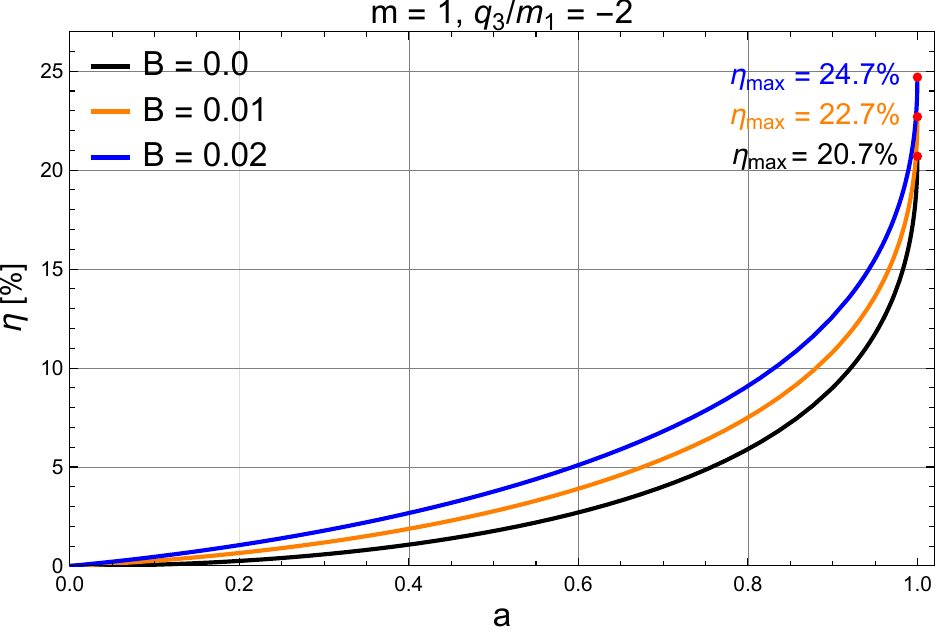}
\includegraphics[scale=0.5]{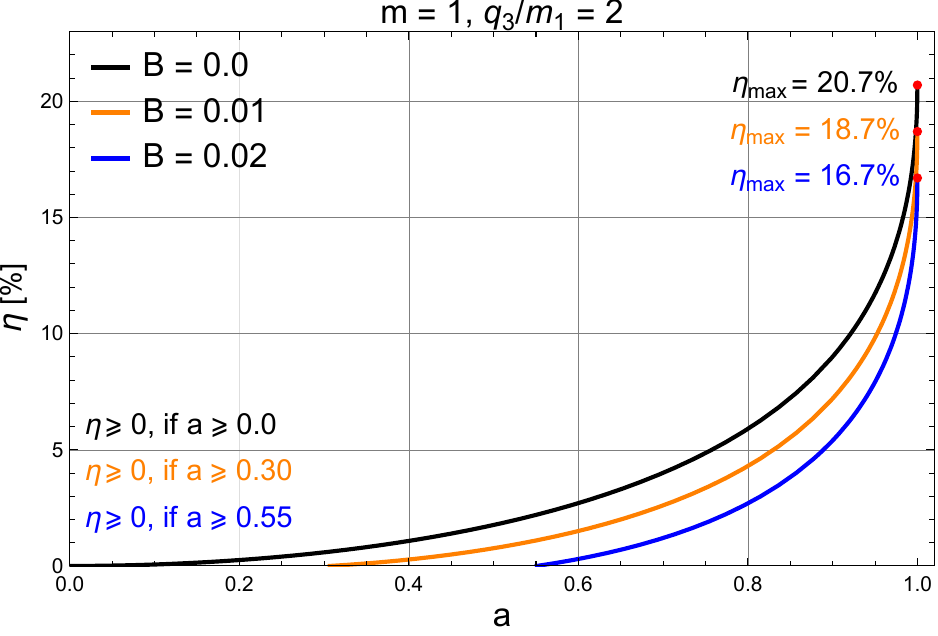}
\includegraphics[scale=0.5]{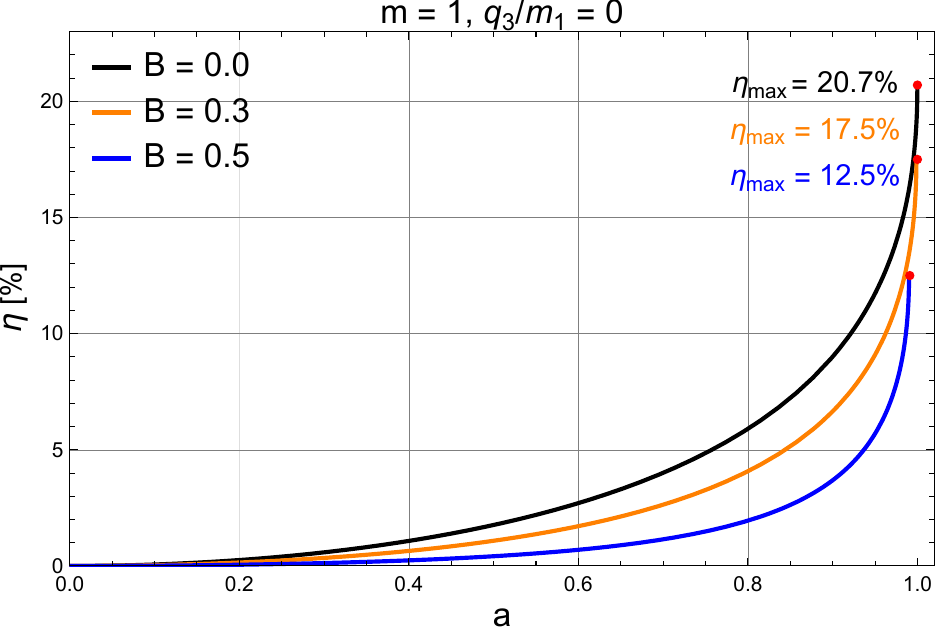}
\includegraphics[scale=0.5]{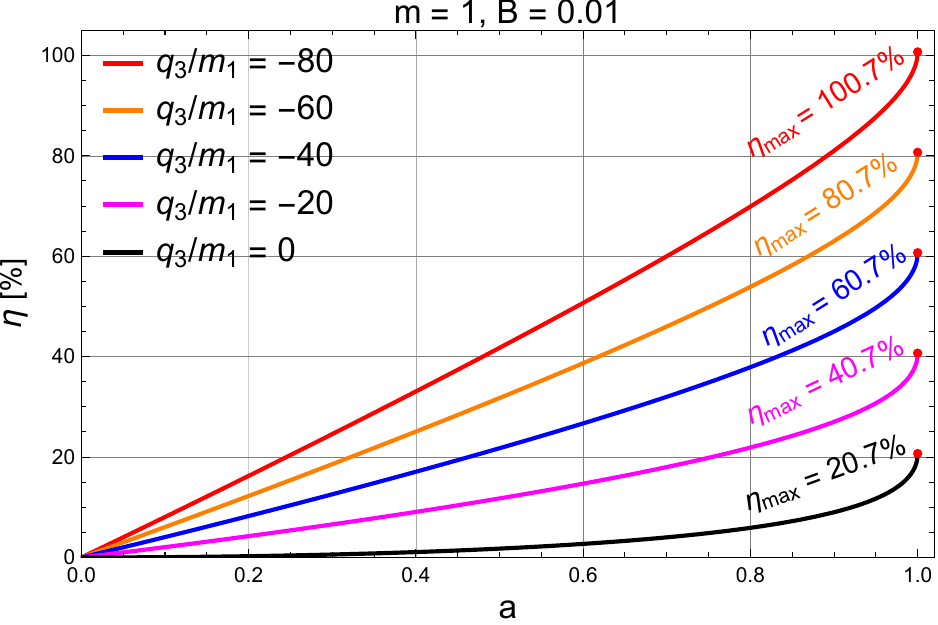}
\caption{\label{fig:en_effKBR} \label{fig:en_eff1} The efficiency of energy extraction from the KBR BH via the MPP as a function of the spin parameter $a$ for different values of magnetic field $B$ and the charge-to-mass ratio parameter $q_3/m_1$. }
\end{figure*}

\begin{figure*}
\includegraphics[scale=0.5]{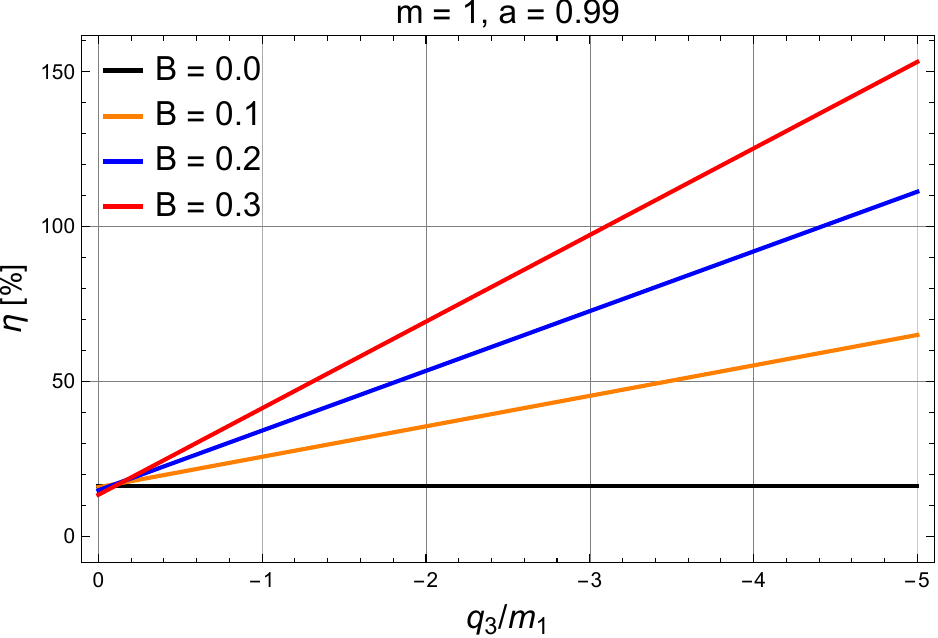}
\includegraphics[scale=0.5]{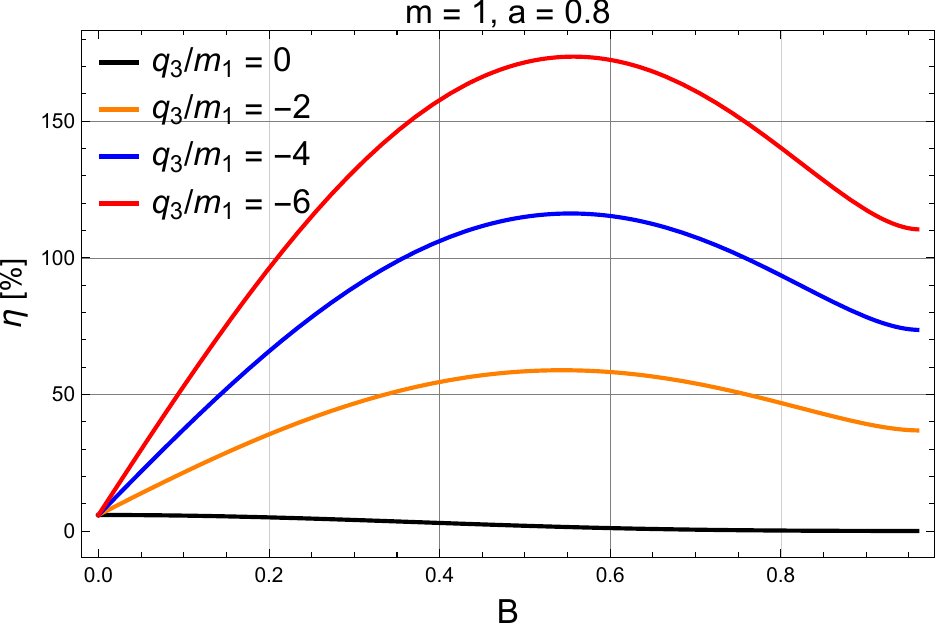}
\caption{\label{fig:en_effKBR} \label{fig:en_eff2} The efficiency of energy extraction from the KBR BH via the MPP as a function of the charge-to-mass ratio parameter $q_3/m_1$ (left panel) and the magnetic field $B$ (right panel). }
\end{figure*}

Fig.~\ref{fig:en_eff_dist} illustrates the distribution of the efficiency of the MPP for decay occurring at different locations around the KBR BH. From the left panel, one can see that for $q_3/m_1 = 0$, the rotational energy can only be extracted inside the ergoregion (the white dashed line). In contrast, for $q_3/m_1 = -2$ (right panel), the energy extraction can be available even outside the ergoregion due to the electromagnetic interactions between the charged particle and the magnetic field. This clearly demonstrates the significant influence of electromagnetic interactions on the efficiency of the process. In both cases, the efficiency reaches its maximum value when the decay occurs closer to the BH and in the equatorial plane.

Fig.~\ref{fig:en_eff1} shows the maximum efficiency as a function of the spin parameter $a$ for various values of the magnetic field $B$ and the charge-mass ratio $q_3/m_1$. For positive values of $q_3/m_1$ (top right), there exists a lower bound on the spin parameter $a$ above which positive efficiency is observed. For zero and negative values of $q_3/m_1$ (top left and bottom left), positive efficiency can be obtained for any positive value of the spin parameter $a$. For negative values of $q_3/m_1$, increasing $B$ to small values also enhances the efficiency. In contrast, for $q_3/m_1 = 0$, increasing $B$ leads to a decrease in the efficiency. For sufficiently large negative values of $q_3/m_1$, the efficiency can exceed $100\%$ (bottom right). Fig.~\ref{fig:en_eff2} shows the maximum efficiency as a function of $B$ and $q_3/m_1$ for fixed values of the spin parameter $a$. As $q_3/m_1$ becomes more negative, the efficiency increases linearly. As $B$ increases, the efficiency initially increases, reaches a maximum value, and then decreases.

\textit{Astrophysical application:} We now explore an astrophysical application of the MPP. Specifically, we estimate the maximum energy of an electron escaping from the ergoregion of the KBR BH. Rather than focusing on extraction efficiency, we calculate the electron energy after it has been accelerated by the magnetic field in the ergoregion. This perspective enables the MPP to be applied to realistic astrophysical scenarios based on the preceding results. To achieve a quantitative description, we investigate neutron beta decay occurring close to the horizon
\cite{PierreAuger:2018qvk,Tursunov:2020juz,2022Symm...14..482T,Khamidov24EPJC...84.1300X}.
\begin{eqnarray}
    n^0 \rightarrow p^+ + W^- \rightarrow p^+ + e^- + \overline{\nu}_e\, .
\end{eqnarray}

Using Eqs.~(\ref{Eq:eta}) and (\ref{En:effmax}), we derive the analytical expression for the energy of electrons accelerating after decay near the horizon in the equatorial plane:
\begin{widetext}
    \begin{align} \label{Eq:E3}
E_{e^-} = \frac{1}{2}\left(\sqrt{\frac{4 m \left(a^2 B^2-1\right)}{r_h \left(a^2 B^2-2\right)}-\frac{4 B^2 m^2 \left(a^2 B^2-1\right)}{\left(a^2 B^2-2\right)^2}-\frac{a^2 B^2}{B^2 r^2_h+1}}-1\right) \times  m_{n^0}c^2 +  \frac{e B a}{\sqrt{1+B^2r_h^2}}\, .
\end{align}
\end{widetext}
We estimate the accelerating power of the supermassive BH at the center of the Milky Way, $\text{SgrA}^*$, with mass
$m = 4.29 \times 10^{6}\,M_{\odot}$~\cite{GravColabM2023}, considering spin parameters
$a \leq 0.1$~\cite{SpinOfSgrA2020ApJ}, $a = 0.52$~\cite{SpinOfSgraA2003Natur}, and $a \approx 0.9$~\cite{SpinOfSgrA2024MNRAS}.
The magnetic field strength on horizon scales is
$B \simeq 1$-$30\,\mathrm{G}$~\cite{MF:2021ApJ}.
We then obtain the minimum and maximum energies of escaping electrons as
\begin{equation}
E_{e^-}^{\mathrm{min}} = 1.89 \times 10^{13}\,\mathrm{eV}
\left( \frac{B}{1\,\mathrm{G}} \right)
\left( \frac{m}{4.29\times10^6\,M_\odot} \right)
\left( \frac{a}{0.1} \right)\, ,
\end{equation}
and
\begin{equation}
E_{e^-}^{\mathrm{max}} = 5.11 \times 10^{15}\,\mathrm{eV}
\left( \frac{B}{30\,\mathrm{G}} \right)
\left( \frac{m}{4.29\times10^6\,M_\odot} \right)
\left( \frac{a}{0.9} \right)\, .
\end{equation}
For comparison, the observed cosmic-ray electron energies span the range
$E_e \simeq 0.3-40\,\mathrm{TeV}$~\cite{NRGofElectrons2024PhRvL}.

\begin{figure} 
\includegraphics[width = 0.48\textwidth]{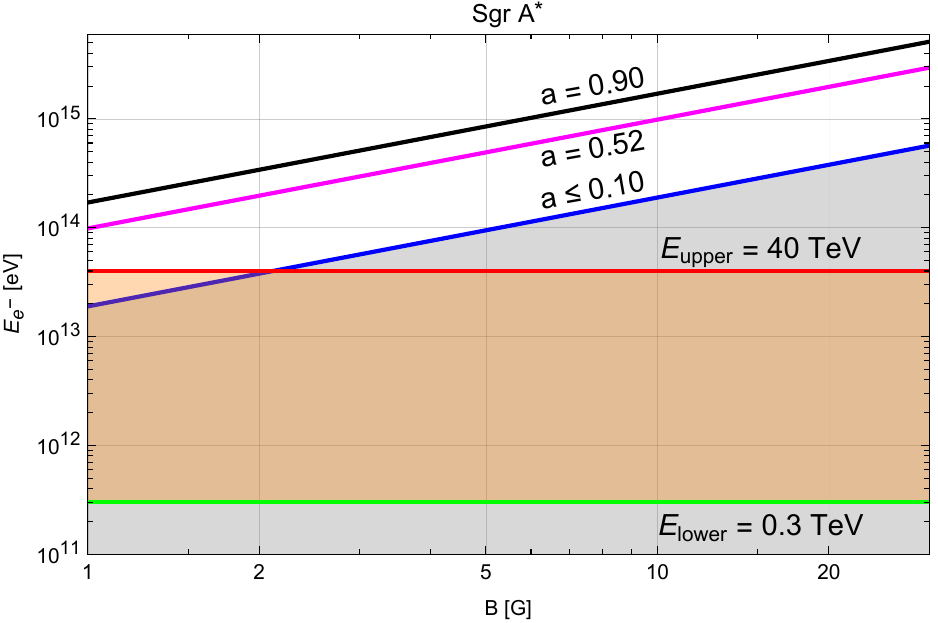}
\caption{ \label{observedElectronEnergy} The energy of accelerated electrons from the KBR BH using the MPP. }
\end{figure}

Fig.~\ref{observedElectronEnergy} illustrates the energies of electrons accelerated by $\text{SgrA}^*$ together with the observed electron energy range in cosmic rays. 
The black, magenta, and blue curves correspond to accelerated electron energies for different values of the spin parameter of $\text{SgrA}^*$. 
The orange shaded region indicates the observed electron energy range. 
The figure demonstrates that $\text{SgrA}^*$ is capable of accelerating electrons to energies within the observed range. 
In particular, $\text{SgrA}^*$ can accelerate electrons up to a maximum energy of
$5.11 \times 10^{15}\,\mathrm{eV}$. 
Although this value lies well above the observed upper limit, electrons can rapidly lose energy through inverse-Compton scattering and synchrotron radiation, causing their energies to fall within the observed cosmic-ray electron range.

\section{Conclusions}
\label{Sec:conclusion}

Astrophysical rotating BHs are widely regarded as the most promising energy sources powering high-energy astrophysical phenomena. This motivates detailed studies of BH energetics, particularly in order to gain a deeper understanding of the rotational energy of magnetized BH systems. In this context, in this paper, we considered an intriguing candidate, the rotating Kerr-Bertotti-Robinson (KBR) BH, which represents an exact solution describing a Kerr BH immersed in a uniform Bertotti-Robinson magnetic field. We begin our analysis by investigating the ergosphere and highlighting the combined effects of the spin parameter $a$ and the magnetic field parameter $B$ on the structure and behavior of the ergoregion. 

We found that the external magnetic field $B$ affects the ergoregion size in a non-monotonic way. For a fixed spin $a$, the ergosphere initially grows as $B$ increases due to enhanced frame dragging due to the magnetically modified coefficients $I_{1}$ and $I_{2}$, but beyond a certain value of $B$ it starts to shrink. This nontrivial behavior is induced by the magnetic field (see Fig.~\ref{fig:Ergosphere}). In addition, we analyzed the behavior of the static limit surface by examining its full angular dependence $r_{\rm SLS}(\theta)$ for different values of the external magnetic field $B$ and the spin parameter $a$. We found that the equatorial radius increases more rapidly than the polar radius, indicating that the ergoregion becomes progressively more oblate as $B$ increases. Moreover, the equatorial extent of the ergoregion displays a non-monotonic behavior, i.e. it initially expands with increasing $B$, reaches a maximum, and then decreases at higher field strengths. However, it should be noted that the ergoregion reaches its largest size at intermediate magnetic field values $B$ (see Fig.~\ref{fig:sls_combined}). 

We then investigated the motion of charged test particles around the rotating KBR BH. Our results show that stable orbits shift toward smaller radial distances as the magnetic field strength $B$ increases. The magnetic field also enhances the effective potential barrier and introduces an additional attractive force toward the BH. Furthermore, we analyzed the ISCO parameters, namely the energy $\mathcal{E}_{\rm ISCO}$, angular momentum $\mathcal{L}_{\rm ISCO}$, and radius $r_{ISCO}$. We found that increasing the spin parameter $a$ reduces the ISCO parameters due to enhanced centrifugal effects. In contrast, the ISCO radius decreases with increasing magnetic field strength $B$, indicating the presence of an additional magnetic field-induced attraction toward the BH.

As noted above, astrophysical rotating BHs are regarded as powerful high-energy sources. In this work, we considered the magnetic Penrose process (MPP) to investigate the efficiency of energy extraction from a rotating KBR BH. We found that, for neutral particles, the efficiency is possible only within the ergoregion-consistent with the Kerr case-and reaches its maximum when the decay occurs in the equatorial plane at radii close to the event horizon. For negatively charged particles, the energy extraction can remain possible even outside the ergoregion due to electromagnetic interactions (see Fig.~\ref{fig:en_eff_dist}). In contrast, for positively charged particles, the efficiency of energy extraction decreases with increasing magnetic field strength $B$ and becomes smaller than that in the Kerr case. Interestingly, the efficiency of energy extraction for negatively charged particles is significantly enhanced by the MPP, reaching values that can exceed 100\% (see Fig.~\ref{fig:en_eff1}). It is important to emphasize that the non-monotonic evolution of the ergoregion with increasing magnetic field $B$ gives rise to a pronounced non-monotonic dependence of the energy extraction efficiency on $B$ (see Fig.~\ref{fig:en_eff2}). Specifically, the efficiency initially increases, reaches a maximum, and then decreases as the extremal condition approaches. This behavior stands in sharp contrast to the monotonic magnetic field effects typically observed in the Kerr geometry.

We further explored an astrophysical application of the MPP by estimating the maximum energy of electrons escaping from the ergoregion of the KBR BH. To model this process, we considered neutron beta decay occurring near the event horizon and derive an analytical expression for the energy gained by electrons accelerated by the magnetic field. Our results indicated that the KBR BH can act as a powerful source of high-energy electrons across a wide range of energies. Applied to $\mathrm{SgrA}^*$, our results showed that electrons can reach energies of $\sim 10^{15}\,\mathrm{eV}$ for realistic spin and magnetic field values. Radiative losses, such as synchrotron and inverse-Compton processes, can subsequently reduce these energies to the observed $\mathrm{TeV}$ range.

Our findings showed that the KBR BH can provide a clear illustration of how intrinsically electromagnetic fields can qualitatively alter BH energetics, leading to an optimal magnetic field strength for rotational energy extraction. In electrovacuum spacetimes where the magnetic field is intrinsic to the geometry, the MPP can be significantly modified with important implications for high-energy astrophysical systems.

\begin{acknowledgements}
S.S. is supported by the National Natural Science Foundation of China under Grant No. W2433018. PS gratefully acknowledges financial support from the Vellore Institute of Technology through its Seed Grant (Grant No.\ SG20230079, 2023). PS and HN also acknowledge the support of the Anusandhan National Research Foundation (ANRF) under the Science and Engineering Research Board (SERB) Core Research Grant (Grant No.\ CRG/2023/008980). 
\end{acknowledgements}

\appendix

\section{ Analytical Behavior of the Static Limit Surface and Event Horizon With Magnetic Field}
\label{app:Ana_ergo_B}

The stationary limit surface (ergosurface) is defined by $g_{tt}=0$, while the event horizon is determined by $g^{rr}=0$, which for the KBR BH metric in Eq.~\eqref{Eq:metric} corresponds to $Q=0$ or equivalently $\Delta=0$ (since $Q=(1+B^2r^2)\Delta$).

\subsection{Event Horizon Analysis}

The event horizon $r_{+}(B)$ satisfies $\Delta(r;B)=0$, where:
\begin{equation}
\Delta = \left( 1 - B^{2}m^{2}\frac{I_{2}}{I_{1}^{2}} \right)r^{2} - 2mr \frac{I_{2}}{I_{1}} + a^{2} = 0.
\label{eq:Delta_exact}
\end{equation}

This is a quadratic equation in $r$:
\begin{equation}
A(B) r^2 - 2C(B) r + a^2 = 0,
\end{equation}

with coefficients:
\begin{align}
A(B) &= 1 - B^{2}m^{2}\frac{I_{2}}{I_{1}^{2}}, \\
C(B) &= m\frac{I_{2}}{I_{1}}.
\end{align}

The exact solution is:
\begin{equation}
r_{\pm}(B) = \frac{C(B) \pm \sqrt{C(B)^2 - a^2 A(B)}}{A(B)}.
\label{eq:horizon_solution}
\end{equation}

For $B=0$, we have $I_1(0)=1$, $I_2(0)=1$, $A(0)=1$, $C(0)=m$, recovering the Kerr horizons:
\begin{equation}
r_{\pm}(0) = m \pm \sqrt{m^2 - a^2}.
\end{equation}

To analyze the behavior with $B$, we compute derivatives. First, note the small-$B$ expansions:
\begin{align}
I_1 &= 1 - \tfrac{1}{2}B^2 a^2 + \mathcal{O}(B^4), \\
I_2 &= 1 - B^2 a^2 + \mathcal{O}(B^4), \\
\frac{I_2}{I_1} &= 1 - \tfrac{1}{2}B^2 a^2 + \mathcal{O}(B^4), \\
\frac{I_2}{I_1^2} &= 1 + \mathcal{O}(B^4).
\end{align}

Thus:
\begin{align}
A(B) &= 1 - B^2 m^2 + \mathcal{O}(B^4), \\
C(B) &= m\left(1 - \tfrac{1}{2}B^2 a^2\right) + \mathcal{O}(B^4).
\end{align}

The outer horizon $r_{+}(B)$, known as the event horizon $r_h (B)$, expands as:
\begin{widetext}
\begin{equation}
r_{h}(B) = \frac{C + \sqrt{C^2 - a^2 A}}{A} = \frac{m\left(1 - \tfrac{1}{2}B^2 a^2\right) + \sqrt{m^2\left(1 - B^2 a^2\right) - a^2\left(1 - B^2 m^2\right) + \mathcal{O}(B^4)}}{1 - B^2 m^2 + \mathcal{O}(B^4)}.
\end{equation}

Simplifying the square root term:
\begin{align}
\sqrt{C^2 - a^2 A} &= \sqrt{m^2\left(1 - B^2 a^2\right) - a^2\left(1 - B^2 m^2\right) + \mathcal{O}(B^4)} \nonumber\\
&= \sqrt{m^2 - a^2} \left(1 + \mathcal{O}(B^4)\right).
\end{align}

Therefore,
\begin{equation}
r_{h}(B) = \frac{m\left(1 - \tfrac{1}{2}B^2 a^2\right) + \sqrt{m^2 - a^2}}{1 - B^2 m^2} + \mathcal{O}(B^4).
\end{equation}

Expanding the denominator:
\begin{equation}
r_{h}(B) = \left[m\left(1 - \tfrac{1}{2}B^2 a^2\right) + \sqrt{m^2 - a^2}\right]\left(1 + B^2 m^2 + \mathcal{O}(B^4)\right).
\end{equation}

Thus to $\mathcal{O}(B^2)$:
\begin{align}
r_{h}(B) &= m + \sqrt{m^2 - a^2} + B^2\left[m^2\left(m + \sqrt{m^2 - a^2}\right) - \tfrac{1}{2}m a^2\right]+ \mathcal{O}(B^4).
\label{eq:rh_B_expansion}
\end{align}

The $B^2$ coefficient is:
\begin{equation}
\frac{d^2 r_{h}}{dB^2}\Big|_{B=0} = 2\left[m^2\left(m + \sqrt{m^2 - a^2}\right) - \tfrac{1}{2}m a^2\right] > 0 \quad \text{for } m > 0.
\end{equation}
\end{widetext}

Since $\frac{dr_{h}}{dB}\big|_{B=0} = 0$ and $\frac{d^2 r_{h}}{dB^2}\big|_{B=0} > 0$, the event horizon \textbf{initially increases} with $B$ for small magnetic fields.

\subsection{Ergosurface Analysis}

The ergosurface $r_{\text{sl}}(B,\theta)$ satisfies $g_{tt}=0$. From Eq.~\eqref{Eq:metric}:

\begin{equation}
g_{tt} = \frac{1}{\omega^2 \rho^2}\left( -Q + P\, a^2 \sin^2\theta \right),
\end{equation}

so the condition is:
\begin{equation}
F(r,\theta;B) \equiv Q(r,B) - P(r,\theta;B)\, a^2 \sin^2\theta = 0.
\label{eq:ergo_condition_exact}
\end{equation}

Substituting the definitions:
\begin{equation}
(1+B^2 r^2)\Delta - \left[1 + B^2\left(m^2\frac{I_2}{I_1^2} - a^2\right)\cos^2\theta\right] a^2 \sin^2\theta = 0.
\label{eq:ergo_eq_exact}
\end{equation}

At $B=0$, this reduces to the Kerr ergosurface:
\begin{eqnarray}
\Delta_{\text{Kerr}} &-& a^2 \sin^2\theta = 0 \nonumber\\ &\Rightarrow& r_{\text{sl}}(0,\theta) = m + \sqrt{m^2 - a^2\cos^2\theta}\, .
\end{eqnarray}

For small $B$, let $r_{\text{sl}}(B,\theta) = r_0(\theta) + B^2 r_2(\theta) + \mathcal{O}(B^4)$ with $r_0(\theta) = m + \sqrt{m^2 - a^2\cos^2\theta}$. Substituting into Eq.~\eqref{eq:ergo_eq_exact} and expanding to $\mathcal{O}(B^2)$:
\begin{align}
&(1+B^2 r_0^2)\left[\Delta_{\text{Kerr}}(r_0) + B^2 \Delta_1(r_0)\right] \nonumber\\
&- \left[1 + B^2\left(m^2 - a^2\right)\cos^2\theta\right] a^2 \sin^2\theta = 0,
\end{align}
where $\Delta_{\text{Kerr}}(r_0) = a^2 \sin^2\theta$ by definition, and $\Delta_1(r)$ is the $\mathcal{O}(B^2)$ correction to $\Delta$.

The $\mathcal{O}(B^2)$ terms give:
\begin{equation}
r_0^2 a^2 \sin^2\theta + \Delta_1(r_0) - \left(m^2 - a^2\right) a^2 \sin^2\theta\cos^2\theta = 0.
\end{equation}

Thus:
\begin{equation}
r_2(\theta) = -\frac{r_0^2 a^2 \sin^2\theta + \Delta_1(r_0) - \left(m^2 - a^2\right) a^2 \sin^2\theta\cos^2\theta}{2(r_0 - m)}.
\end{equation}

The sign of $r_2(\theta)$ determines whether the ergosurface initially increases ($r_2>0$) or decreases ($r_2<0$). This is angle-dependent and parameter-dependent.

\subsection{Ergoregion Thickness}

Define the ergoregion thickness as $\Delta r_{\text{ergo}}(B,\theta) = r_{\text{sl}}(B,\theta) - r_{h}(B)$. From Eq.~\eqref{eq:rh_B_expansion}:
\begin{equation}
r_{h}(B) = r_{h}(0) + B^2 \alpha_{h} + \mathcal{O}(B^4),
\end{equation}

with $\alpha_{h} = m^2\left(m + \sqrt{m^2 - a^2}\right) - \tfrac{1}{2}m a^2 > 0$.

The ergosurface expansion is:
\begin{equation}
r_{\text{sl}}(B,\theta) = r_{\text{sl}}(0,\theta) + B^2 r_2(\theta) + \mathcal{O}(B^4).
\end{equation}

Therefore:
\begin{equation}
\Delta r_{\text{ergo}}(B,\theta) = \left[r_{\text{sl}}(0,\theta) - r_{h}(0)\right] + B^2\left[r_2(\theta) - \alpha_{h}\right] + \mathcal{O}(B^4).
\end{equation}

Since $r_{\text{sl}}(0,\theta) > r_{h}(0)$ for $\theta \neq 0,\pi$, the unperturbed thickness is positive. The $B^2$ correction term $r_2(\theta) - \alpha_{h}$ determines whether the ergoregion initially expands or contracts.

At the equator ($\theta=\pi/2$), $r_{\text{sl}}(0,\pi/2)=2m$, and:
\begin{align}
r_2(\pi/2) &= -\frac{(2m)^2 a^2 + \Delta_1(2m) - 0}{2(2m - m)} \nonumber\\
&= -\frac{4m^2 a^2 + \Delta_1(2m)}{2m}.
\end{align}

Typically, it $\Delta_1(r)$ is negative (since magnetic fields tend to contract the geometry), making $r_2(\pi/2) < 0$. Thus $r_2(\pi/2) - \alpha_{h}$ is negative, meaning the ergoregion thickness decreases at the equator.

However, at other angles, $r_2(\theta)$ could be positive if:
\begin{equation}
\left(m^2 - a^2\right) a^2 \sin^2\theta\cos^2\theta > r_0^2 a^2 \sin^2\theta + \Delta_1(r_0).
\end{equation}

This could occur near the poles ($\theta \approx 0$) where the right side is small.

\subsection{Large-$B$ Behavior}

For $B \gg 1$, the coefficients scale as:
\begin{align}
I_1 &\sim -\tfrac{1}{2}B^2 a^2, \\
I_2 &\sim -B^2 a^2, \\
\frac{I_2}{I_1} &\sim 2, \\
\frac{I_2}{I_1^2} &\sim -\frac{4}{B^2 a^2}.
\end{align}

Thus:
\begin{align}
A(B) &\sim 1 + 4m^2 + \mathcal{O}(B^{-2}), \\
C(B) &\sim 2m.
\end{align}

The horizon equation becomes:
\begin{equation}
(1+4m^2)r^2 - 4m r + a^2 = 0,
\end{equation}

with positive discriminant for sufficiently large $m$. The horizon approaches a finite limit:
\begin{equation}
\lim_{B\to\infty} r_{h}(B) = \frac{2m \pm \sqrt{4m^2 - a^2(1+4m^2)}}{1+4m^2}.
\end{equation}

For the ergosurface, in the large-$B$ limit, Eq.~\eqref{eq:ergo_eq_exact} becomes dominated by $B^2$ terms:
\begin{equation}
B^2 r^2 \Delta \sim B^2\left(m^2\frac{I_2}{I_1^2} - a^2\right) a^2 \sin^2\theta\cos^2\theta.
\end{equation}

Since $\frac{I_2}{I_1^2} \sim -\frac{4}{B^2 a^2}$, the right side approaches a constant, while the left side grows with $r^2$. This forces $r_{\text{sl}}$ to approach $r_{h}$ where $\Delta=0$.

In summary: (i) Both $r_{h}$ and $r_{\text{sl}}$ can initially increase for small $B$, with $r_{h}$ definitely increasing due to a positive second derivative at $B=0$. (ii) The ergoregion thickness $\Delta r_{\text{ergo}}$ generally decreases with $B$, as the contraction of the gap between the surfaces outweighs their individual movements. (iii) At large $B$, both surfaces approach finite limits with $r_{\text{sl}}\to r_{h}$, causing the ergoregion to vanish. This explains the non-monotonic behavior: individual surfaces may rise then fall, but the region between them consistently shrinks under increasing magnetic influence.

\bibliography{Ref,NewRef}

\end{document}